\documentclass[]{aastex631}

\usepackage{color}
\usepackage{amsmath}
\usepackage{amssymb}
\usepackage{mathrsfs}
\usepackage{gensymb}

\begin{document}

\title{A young white dwarf orbiting PSR~J1835$-$3259B in the bulge globular cluster NGC~6652}

\author[0000-0002-8004-549X]{Jianxing Chen}
\affiliation{Dipartimento di Fisica e Astronomia ``Augusto Righi'',
  Alma Mater Studiorum Universit\`a di Bologna, via Piero Gobetti
  93/2, I-40129 Bologna, Italy}
\affiliation{INAF-Osservatorio di Astrofisica e Scienze dello Spazio
  di Bologna, Via Piero Gobetti 93/3 I-40129 Bologna, Italy}

\author[0000-0002-5038-3914]{Mario Cadelano} 
\affiliation{Dipartimento
  di Fisica e Astronomia ``Augusto Righi'', Alma Mater Studiorum
  Universit\`a di Bologna, via Piero Gobetti 93/2, I-40129 Bologna,
  Italy} 
\affiliation{INAF-Osservatorio di Astrofisica e Scienze dello
  Spazio di Bologna, Via Piero Gobetti 93/3 I-40129 Bologna, Italy}

\author[0000-0002-7104-2107]{Cristina Pallanca}\affiliation{Dipartimento di Fisica e Astronomia ``Augusto Righi'',
  Alma Mater Studiorum Universit\`a di Bologna, via Piero Gobetti
  93/2, I-40129 Bologna, Italy}
\affiliation{INAF-Osservatorio di Astrofisica e Scienze dello Spazio
  di Bologna, Via Piero Gobetti 93/3 I-40129 Bologna, Italy}

\author[0000-0002-2165-8528]{Francesco R. Ferraro}
\affiliation{Dipartimento di Fisica e Astronomia ``Augusto Righi'',
  Alma Mater Studiorum Universit\`a di Bologna, via Piero Gobetti
  93/2, I-40129 Bologna, Italy}
\affiliation{INAF-Osservatorio di Astrofisica e Scienze dello Spazio
  di Bologna, Via Piero Gobetti 93/3 I-40129 Bologna, Italy}

\author[0000-0001-5613-4938]{Barbara Lanzoni}
\affiliation{Dipartimento di Fisica e Astronomia ``Augusto Righi'',
  Alma Mater Studiorum Universit\`a di Bologna, via Piero Gobetti
  93/2, I-40129 Bologna, Italy}
\affiliation{INAF-Osservatorio di Astrofisica e Scienze dello Spazio
  di Bologna, Via Piero Gobetti 93/3 I-40129 Bologna, Italy}

\author[0000-0002-8811-8171]{Alina G. Istrate}
\affil{Department of Astrophysics/IMAPP, Radboud University Nijmegen, PO Box 9010, NL-6500 GL Nijmegen, the Netherlands}

\author[0000-0002-8265-4344]{Marta Burgay}
\affiliation{INAF - Osservatorio Astronomico di Cagliari, via della Scienza 5, I-09047 Selargius (CA), Italy}

\author[0000-0003-1307-9435]{Paulo C. C. Freire}
\affiliation{Max-Planck-Institut für Radioastronomie MPIfR, Auf dem Hügel 69, D-53121 Bonn, Germany}

\author[0000-0002-8396-2197]{Tasha Gautam}
\affiliation{National Radio Astronomy Observatory, 520 Edgemont Rd., Charlottesville, VA 22903, USA}

\author[0000-0001-5902-3731]{Andrea Possenti}
\affiliation{INAF - Osservatorio Astronomico di Cagliari, via della Scienza 5, I-09047 Selargius (CA), Italy}

\author[0000-0001-6762-2638]{Alessandro Ridolfi}
\affiliation{INAF - Osservatorio Astronomico di Cagliari, via della Scienza 5, I-09047 Selargius (CA), Italy}
\affiliation{Max-Planck-Institut für Radioastronomie MPIfR, Auf dem Hügel 69, D-53121 Bonn, Germany}


\begin{abstract}
We report on the discovery of the companion star to the millisecond pulsar PSR J1835-3259B in the
Galactic globular cluster NGC 6652. Taking advantage of deep photometric archival observations acquired through the \textit{Hubble Space Telescope} in near-ultraviolet and optical bands, we identified a bright and blue object at a position compatible with that of the radio pulsar. The companion is located along the helium-core white dwarf cooling sequence and the comparison with binary evolution models provides a mass of $0.17 \pm 0.02~M_\odot$, a surface temperature of $11500\pm1900$~K and a very young cooling age of only $200\pm100$~Myr. The mass and the age of the companion are compatible with a progenitor star of about $0.87~M_{\odot}$, which started transferring mass to the primary during its evolution along the sub-giant branch and stopped during the early red giant branch phase. Combining together the pulsar mass function and the companion mass, we found that this system is observed at an almost edge-on orbit and hosts a neutron star with a mass of $1.44 \pm 0.06~M_\odot$, thus suggesting a highly non-conservative mass accretion phase.
{The young age of the WD companion is consistent with the scenario of a powerful, relatively young MSP indicated by the earlier detection of gamma-rays from this system.}

\end{abstract}


\section{Introduction} \label{sec:intro}

Millisecond pulsars (MSPs) are rapidly spinning neutron stars (NSs) with periods less than 30~ms \citep{Lyne+1998} that are distinguished from the ``normal" pulsars population \citep{Lyne+1998}, which have longer spin periods. Since the first MSP was discovered in the globular cluster (GC) M28 in 1987 \citep{Lyne+1987}, further 278 MSPs have been discovered in 38 Galactic GCs so far\footnote{Pulsars in GCs: \url{http://www.naic.edu/~pfreire/GCpsr.html}}. More than half of these MSPs are located in a binary system. Although  $\sim600$ MSPs have been discovered in the Milky Way\footnote{The ATNF pulsar catalog: \url{https://www.atnf.csiro.au/research/pulsar/psrcat/}}, more than one-third of them are hosted by GCs, thus suggesting a link between the MSP formation rate and the environment in which are formed. In fact, GCs are collisional systems where the high stellar densities and frequent dynamic interactions favor the formation of a large number of exotic objects such as MSPs \citep[e.g.][]{Ransom+2005, Freire+2017, Cadelano+2018,  Ridolfi+2022}, blue straggler stars \citep[e.g.][]{Ferraro+1993, Ferraro+1997a, Ferraro+1999, Ferraro+2004, Cadelano+2022} and cataclysmic variables \citep[e.g.][]{Rivera+2018, Belloni+2019, Belloni+2020}. Therefore, GCs provide the ideal stellar laboratories to study the formation and evolution of exotic systems,  as well as the complex interplay between stellar evolution and dynamics \citep{Ferraro+2009, Ferraro+2012, Ferraro+2018, Ferraro+2019, Ferraro+2020, Dalessandro+2013a, Cadelano+2017a, Lanzoni+2019, Libralato+2022}.

MSPs are generally considered the outcome of the evolution of low-mass X-ray binaries (LMXBs) with a NS primary
(see \citet{Bhatta+1991,Wijnands+1998,Ferraro+2015}). According to this scenario, a slowly rotating NS is spun-up through mass accretion and angular momentum transfer from an evolving companion star. At the end of a $\lesssim1$ Gyr long accretion phase, an exhausted and envelope stripped companion star is left, commonly a He-core white dwarf (WD) \citep{Antoniadis+2016a, Cadelano+2015b, Driebe+1998, Ferraro+2003a, Istrate+2014, Tauris+1999}. A different class of binary MSPs in tight orbits with Period $<$ 1~d are called ``Spider" MSPs \citep{Roberts+2013}. They are characterized by having  non-degenerate companion stars which are usually tidally distorted and heated by the interaction with the pulsar and its relativistic wind \citep{Douglas+2022}. The formation of MSPs can be also explained through alternative channels such as the Accretion-Induced Collapse (AIC) of a massive  Oxygen-Neon or Oxygen-Neon-Magnesium WD. These systems should be characterized by large orbital periods  \citep{Nomoto+1987, Tauris+2013, Freire+2014, Tauris+2017, Ablimit+2022, Wang+2022}. 

In most cases, MSPs are expected to form in the LMXB channel. Mass accretion during this phase can in principle produce NSs with masses higher than standard slowly rotating pulsars. Therefore, MSPs studies can provide a tool to determine through observations the NS mass distribution and maximum mass sustainable by a NS, which is one of the most valuable parameter to constrain the equation of state of ultra-dense matter \citep[see][]{Lattimer+2012, Antoniadis+2016a, Ozel+2016}.

We are leading a long-term program aimed at identifying MSPs in GCs through near-UV to near-IR photometric and spectroscopic observations of their companion stars with the aim of obtaining a complementary view of the binary properties and studying their formation and evolution in the ideal GC laboratory. This program led to the discovery and characterization of several He-core WD companions \citep{Ferraro+2003a, Cadelano+2015b, Cadelano+2019}, one of them possibly orbiting a high-mass NS \citep{Cadelano+2020}, and several companions to spider MSPs \citep{Ferraro+2001a, Sabbi+2003a, Sabbi+2003b, Mucciarelli+2013, Cocozza+2006, Pallanca+2010, Pallanca+2013, Pallanca+2014, Cadelano+2015a, Cadelano+2017b}.

Here we report on the identification and characterization of the companion star to NGC~6652B based on high-resolution near-ultraviolet and optical observations obtained from the \textit{Hubble Space Telescope} (HST).

NGC 6652 is a GC located around 10~kpc from the Sun \citep[][2010 edition]{Harris+1996}, with an age of 13.25$\pm$0.5~Gyr \citep{Dotter+2010}, an intermediate metallicity of [Fe/H] $\approx$ -0.75  and a small extinction of E(B-V) = 0.09 \citep{Harris+1996}(2010 edition) for a stellar system  located in the Galactic bulge. PSR~J1835$-$3259A was the first binary MSP discovered in the cluster by \citet{DeCesar+2015} and it is characterized by a pulse period of 3.89 ms, an orbital period of 9.25 days and a very high eccentricity $e=0.97$. The MSP is orbiting a companion with a minimum mass of about $0.7~M_{\odot}$, which could be either a massive white dwarf or a secondary NS. This system is likely formed through an exchange encounter in the dense cluster environment. Recently, a new binary MSP, namely J1835$-$3259B (hereafter NGC~6652B) has been discovered by \citet{Gautam+2022}. The NS is spinning at 1.83 ms and is located in a 1.2 days long orbit with an extremely small eccentricity of about $3\times10^{-5}$. \citet{Zhang+2022} used observations obtained by the Fermi Gamma-ray Space Telescope and detected a high-energy emission of the MSP as a source with a $\gamma$-ray luminosity of $5\times 10^{34} \, \rm{erg\ s^{-1}}$.

This paper is organized as follows: in section~\ref{sec:2}, we describe the data set and data reduction; the identification and characterization of the companion star are presented in section ~\ref{sec:3} and the conclusions are finally drawn in section~\ref{sec:4}.

\section{Data Reduction}
\label{sec:2}
In this work, we used high-resolution and deep photometric data acquired with the HST in the near-ultraviolet band, captured by the UVIS channel of the Wide Field Camera 3 (WFC3), the obtained archival data set is a part of the HST legacy survey of galactic GCs under GO-13297 (PI: Piotto). There are 3 different filters used: F275W(near-UV), F336W(U), and F438W(B), the observational log of the data set is listed in Table~\ref{table01}. 

\begin{table}
\centering
\caption{Observation Log of the data set.}
\label{table01}
\begin{tabular}{cccc}
\hline
\hline
obs. ID       & F275W    & F336W  & F438W \\ \hline
	  		  & 690s$\times$2 	 & 305s$\times$2 & 60s$\times$1 \\
GO-13297      & 775s$\times$2    & 313s$\times$1 & 69s$\times$1 \\
              & 800s$\times$2    & 313s$\times$3 & 86s$\times$1 \\
\hline
\end{tabular}
\end{table}


The data reduction was performed on the calibrated images with extension $\_flc$ 
(UVIS calibrated exposure including charge transfer efficiency correction), after pre-reduction including Pixel Area Map correction.

In searching for the optical companion to PSR~J1835$-$3259A and PSR~J1835$-$3259B we adopted the so-called ``UV-route", a photometric procedure specifically optimized for the 
identification of blue and hot objects in crowded fields such as hot Horizontal Branch stars, BSSs and WDs and widely used by our group in previous papers (see \citealt{FP+1993,Ferraro+1997a,Ferraro+1997b,Ferraro+2001b, Ferraro+2003b, Dalessandro+2013b}). Specifically, in this paper we followed the approach described in \citet{Cadelano+2019,Cadelano+2020} (see also \citealt{Raso+2017, Raso+2020, Chen+2021, Chen+2022}).

As the first step, about 250 bright and unsaturated stars were selected to model the point-spread function (PSF) for each image, then the resulting models were applied to all the sources detected above $4~\sigma$ from the background. As a second step, we created a ``\textit{master list}" of stars containing all the sources detected in at least half the F275W images. The PSF-fitting of all the sources in the master-list was forced at the corresponding positions in all the frames using  DAOPHOT/ALLFRAME \citep{Stetson+1994}. Finally, magnitudes obtained for different stars were averaged and homogenized using DAOPHOT/DAOMASTER.

The instrumental magnitudes were calibrated to the VEGAMAG system by using appropriate zero-points and aperture corrections\footnote{\url{https://www.stsci.edu/hst/instrumentation/wfc3/data-analysis/photometric-calibration}}. After correction for geometric distortions \citep{Bellini+2011}, the coordinates were aligned to the International Celestial Reference System by cross-correlation with the Gaia DR3 Catalog \citep{Gaia+2022}. To this aim, we used the cross-correlation software CataXcorr\footnote{\url{http://davide2.bo.astro.it/~paolo/Main/CataPack.html}} adopting a six parameter linear transformation to convert the instrumental (x,y) positions to the (RA,Dec) absolute coordinate system. The transformation residuals returned a combined astrometric uncertainty of about 14 mas.

\section{The companion to NGC 6652B}
\label{sec:3}

\subsection{Identification of the companion star}

\begin{figure}
    \centering
    \includegraphics[width=5.6cm]{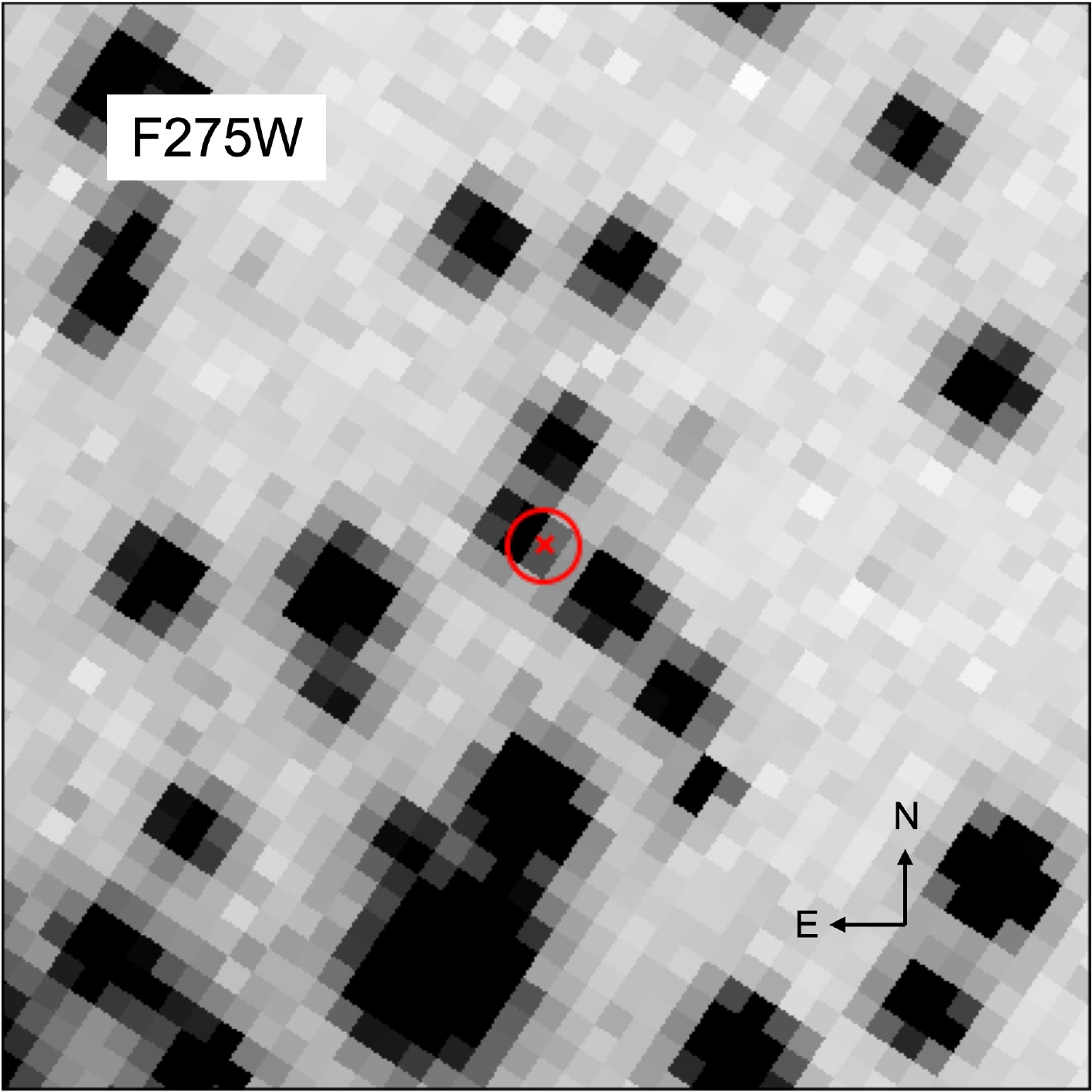}
    \hspace{5pt}
    \includegraphics[width=5.6cm]{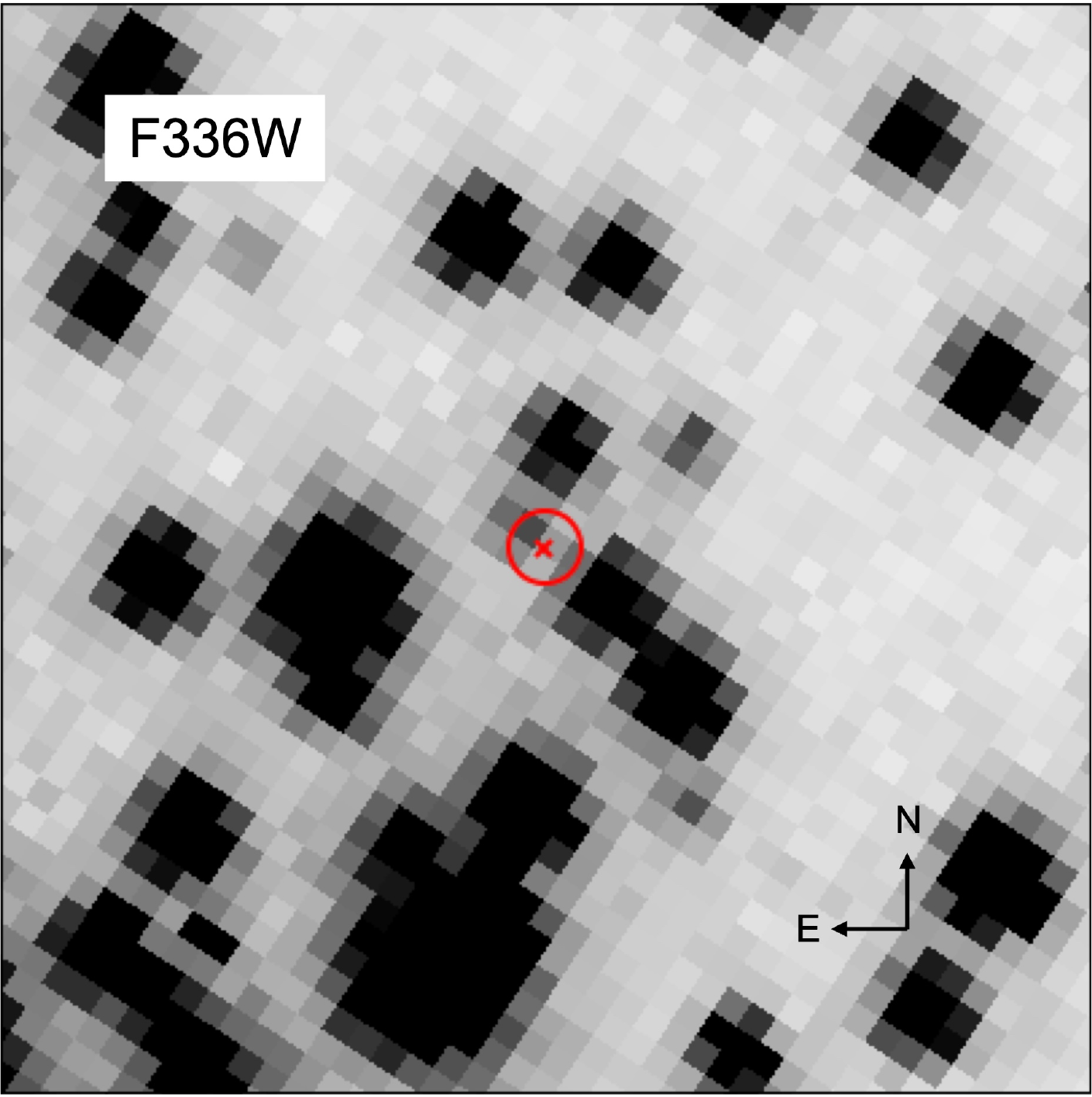}
    \hspace{5pt}
    \includegraphics[width=5.6cm]{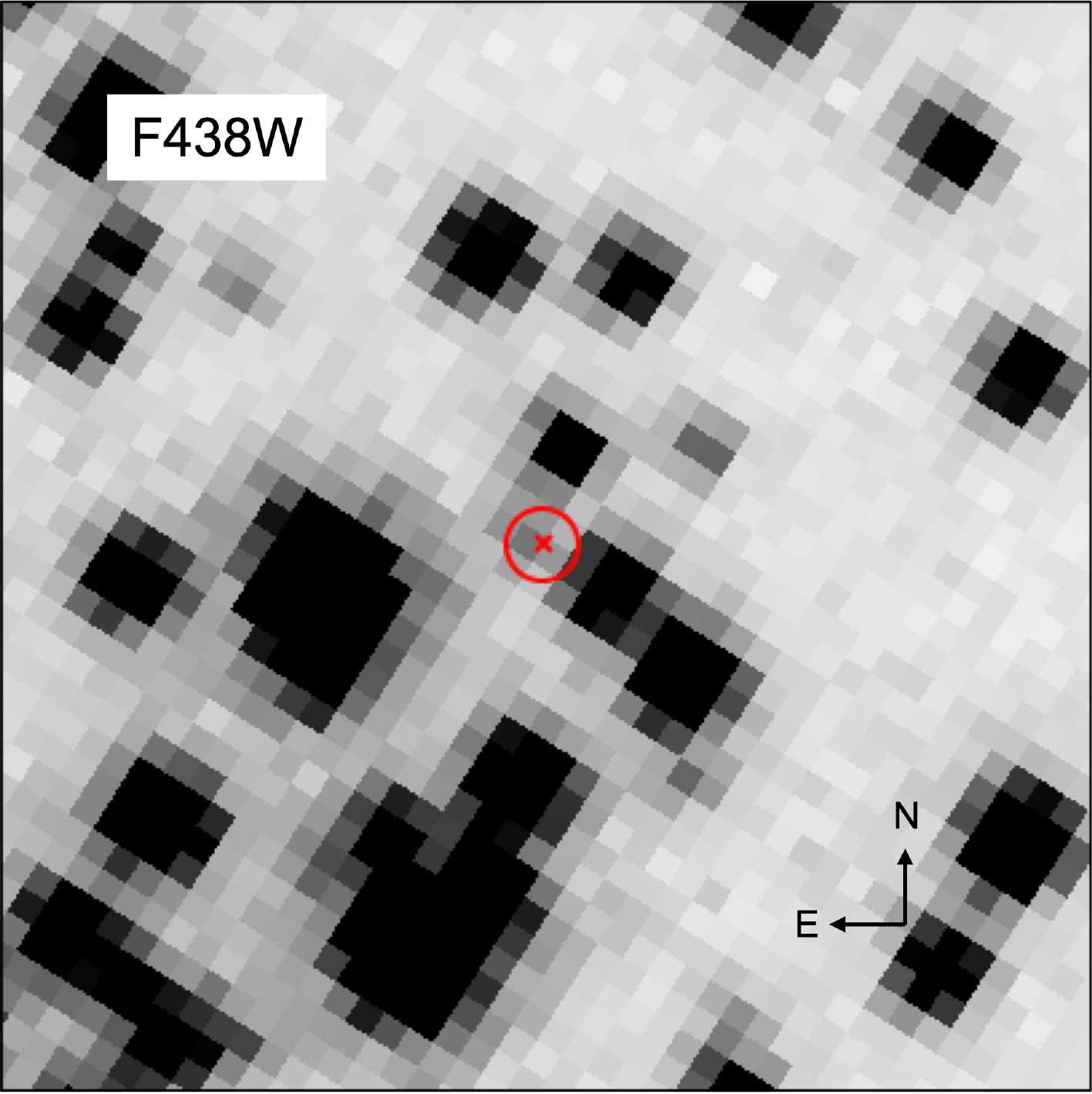}
    \caption{$1.5\arcsec \times 1.5\arcsec$ finding chart of the NGC~6652B companion in the F275W, F336W, and F438W filters. The red circle has a radius of 50 mas ($\sim3\sigma$ uncertainty), while the center of the circle corresponds to the position of the MSP. It is clear that the candidate counterpart (a white dwarf) gets fainter in redder filters, while most of the other stars get brighter. }
    \label{fig:1}
\end{figure}

To identify the optical counterpart of the MSPs in the cluster, we carefully analyzed all the stars in the region surrounding the two MSP positions. No stars are found at a position compatible with PSR~J1835$-$3259A and we provide the finding chart and magnitude upper limits for this object in the Appendix~\ref{sec:app}. On the other hand, a peculiar bright and blue star is found at 30 mas from the timing position of NGC~6652B. The resulting $1\sigma$ radio timing and optical uncertainty for this MSP is 17~mas (see Section~\ref{sec:2} and Table~3 in \citet{Gautam+2022}), therefore the two positions of the two sources turn out to be compatible within  $1.8 \sigma$. Figure~\ref{fig:1} shows the finding chart in the three different adopted filters. It can be easily appreciated that the source closest to NGC~6652B is a very blue star as its luminosity rapidly decreases in redder filters. Nonetheless, the candidate is bright enough to be detected in all the three filters with small uncertainties: $m_{F275W} = 21.77 \pm 0.02$, $m_{F336W} = 21.93 \pm 0.04$, and $m_{F438W} = 22.62 \pm 0.07$. Observations in the F606W and F814W filters are available in the HST archive but we did not include them in the analysis since the companion is not detectable due to a nearby heavily saturated star which totally hampers ant meaningful estimate of the candidate magnitudes in these filters.

\begin{figure}
	\centering
	\includegraphics[width=14cm]{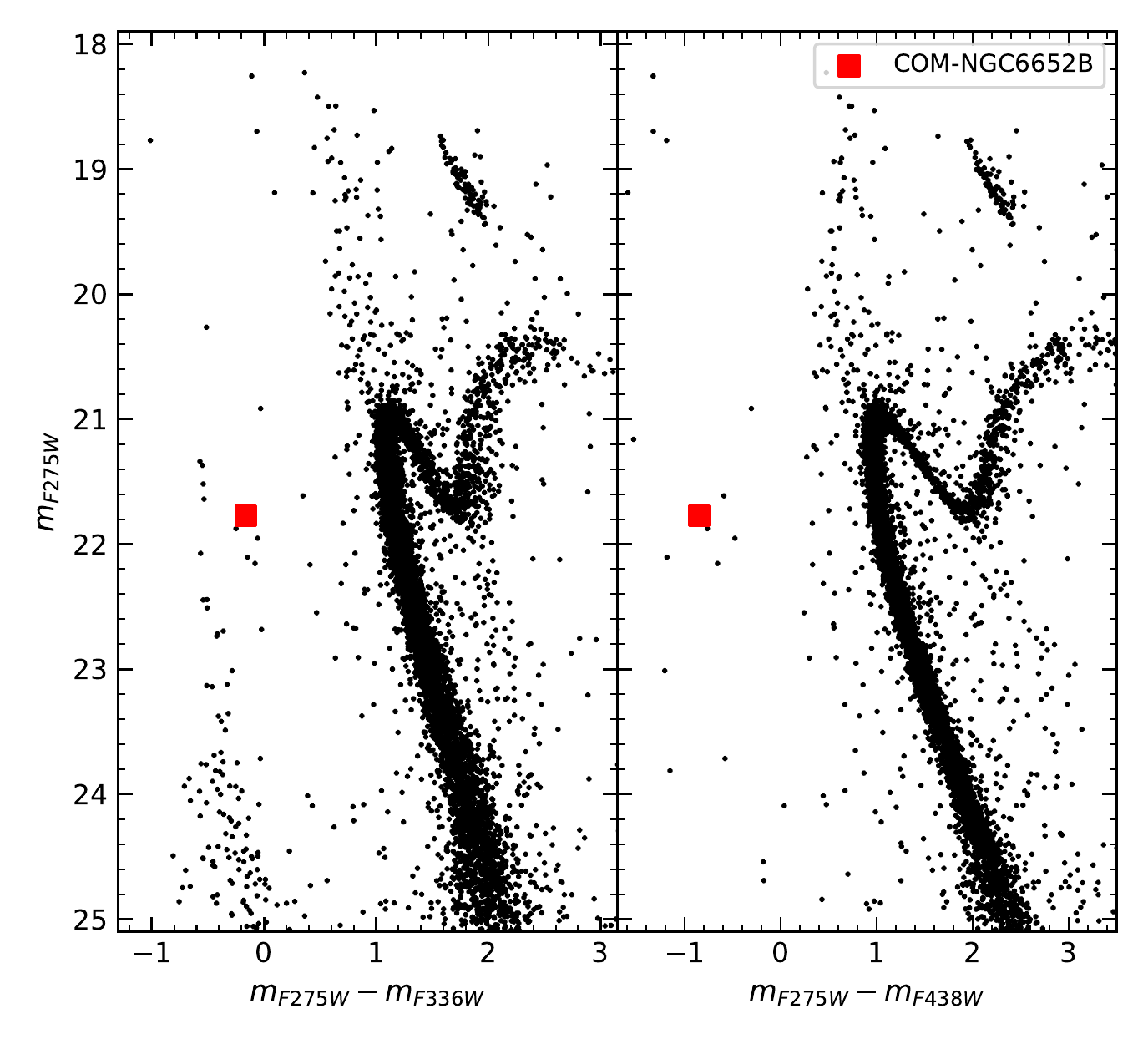}
	\caption{CMDs of NGC~6652. Left-panel and right-panel are ($m_{F275W}-m_{F336W}, m_{F275W}$), and ($m_{F275W}-m_{F438W}, m_{F275W}$) filter combinations, respectively. The He-core WD companion to NGC~6652B is marked with a red square in each panel. The error bars are smaller than the red squares.}
	\label{fig:2}
\end{figure}

Figure~\ref{fig:2} shows the color-magnitude diagram (CMD) of the cluster in two different filter combinations. Indeed, in both the color combinations shown in Figure~\ref{fig:2}, the candidate companion star (highlighted with a large red square) is located in the CMD region between the Main Sequence and the CO-WD cooling sequence (which is clearly visible in the left panel at color $m_{F275W}-m_{F336W}<0$). Considering the photometric errors, its position turns out to be incompatible (at more than $10\sigma$) with both the MS and the CO-WD cooling sequence. On the other hand, this is the portion of the CMD, where the He-WD cooling sequences are expected to lie. These WDs are the standard outcome of the mass-transfer processes and typically found to be orbiting MSPs. Moreover, due to its luminosity we can also expect that this is a relatively young WD. Thus on the basis of its peculiar position in the CMD we can tentatively hypothesize that the  companion to NGC~6652B is a recently formed He-WD.

The adopted data-set is composed of multi-epoch observations. We investigated the magnitude of the counterpart in the different frames but no evidence of photometric variability could be detected for the companion star. While this can be due to the poor sampling of the binary orbit, He-WD companions only rarely show variability, which is usually not related to the binary orbit but to stellar pulsations (e.g., \citealt{Maxted+2013, Kilic+2015, Antoniadis+2016a,  Parsons+2020}). Indeed, heating and/or tidal distortions are negligible in the case of a WD companion (although exceptions exist, see, e.g., \citealt{Edmonds+2001, Kaplan+2012}), while they become significant in the case of non-degenerate and Roche-Lobe filling companions (e.g., \citealt{Ferraro+2001a, Pallanca+2010, Pallanca+2014, Cadelano+2015a}).

\subsection{Comparison with Binary evolution models}

\begin{figure}
    \centering
    \includegraphics[width=12.5cm]{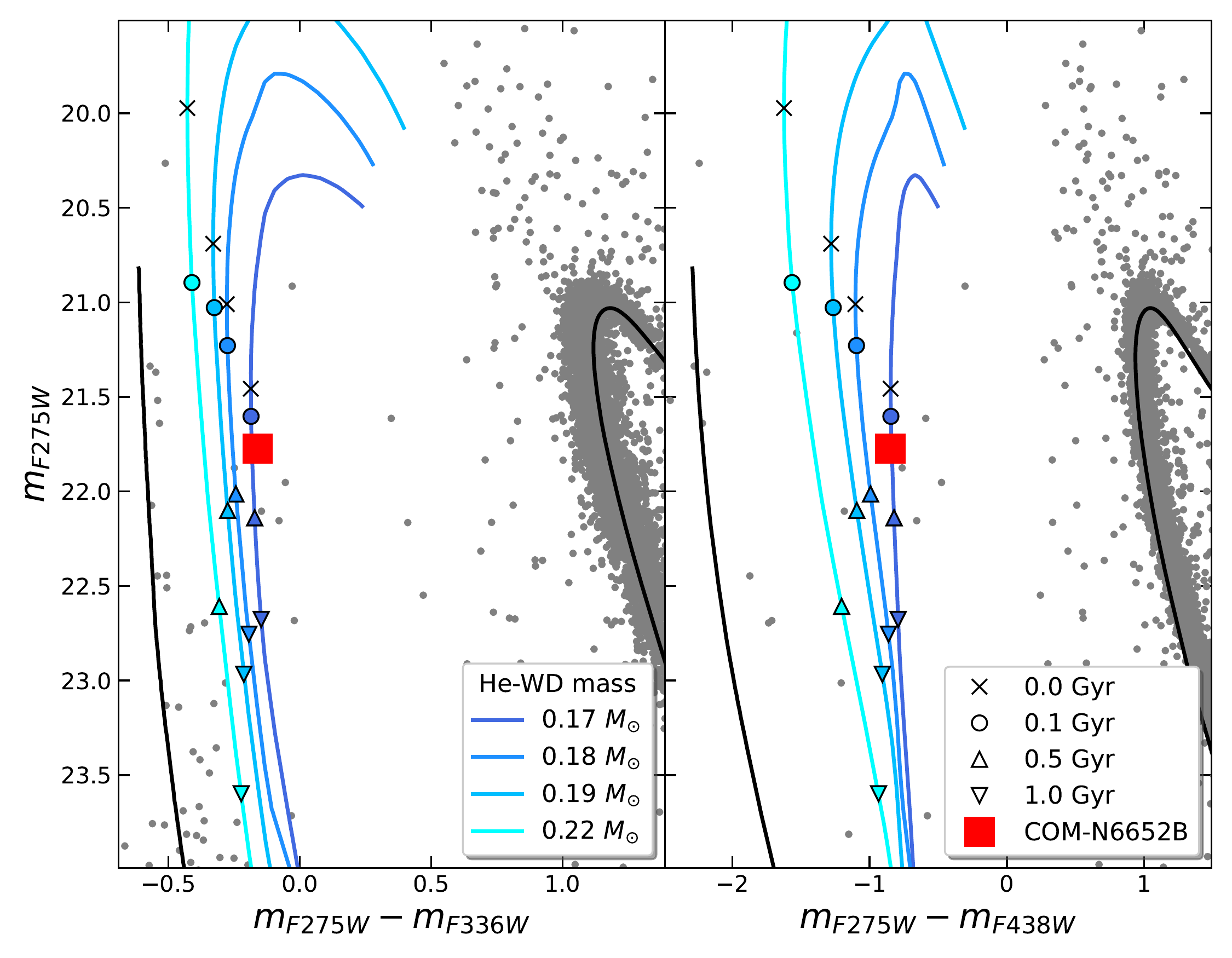}
    \caption{Same CMD as figure~\ref{fig:2}, but the WD region is zoomed in. A theoretical cooling track of a $0.55~M_\odot$ CO-WD and an isochrone of 13.25~Gyr stellar population are marked with black lines, respectively. The other curves in color are cooling tracks of He-WD with masses of $0.17M_\odot$, $0.18M_\odot$, $0.19M_\odot$,$0.22M_\odot$, from left to right, where the cooling age are marked with different symbols in tracks as reported in legend.}
    \label{fig:3}
\end{figure}

In order to confirm the hypothesis that the companion to NGC~6652B is a He-WD, and to derive its physical properties, we compared its position in the CMD with the prediction of theoretical binary evolution models. As first step, we performed a calibration sanity-check by comparing the observed standard evolutionary sequences (such as main-sequence and red giants) with an isochrone extracted from BaSTI database \citep{Hidalgo+2018, Pietrinferni+2021} with an age of 13.25~Gyr \citep{Dotter+2010}, a metallicity [Fe/H] = -0.8 \citep[][2010 version]{Harris+1996}, and [$\alpha$/Fe] = +0.4. We also  compared the position of the observed WD cooling sequence with a theoretical cooling track extracted from BaSTI database \citep{Salaris+2022} for CO-WDs with a canonical mass of $0.55~M_\odot$. Absolute magnitudes were converted to the observed frame by adopting a distance modulus of $(m-M)_0 = 14.97$ and a color excess of $E(B-V) = 0.1$ (in very good agreement with the values quoted by \citealt{Harris+1996}) and using appropriate extinction coefficients from \citet{Cardelli+1989, ODonnell+1994}. As shown in Figure~\ref{fig:3}, both the isochrone and the CO-WD cooling track nicely reproduce the observed evolutionary sequences, thus confirming that the photometric calibration and cluster parameters are sufficiently precise to allow the next step of exploration. 

In order to constrain the properties of the companion star, we extracted binary evolution models from the database described in \citet{Istrate+2014, Istrate+2016}. These models follow the evolution of a NS binary during the whole mass-transfer stage, the proto-WD and the WD cooling stage. 
 The resulting evolution tracks cover a large parameter space with a He-WD mass range of $0.17~M_\odot - 0.4~M_\odot$, a surface temperature of 5000 K - 20000 K, and cooling ages\footnote{Following \citet{Istrate+2016}, the WD cooling age is defined as the time passed since the proto-WD reached the maximum surface temperature along the evolutionary track.} down to the cluster age. By using the Astrolib Pysynphot Package \footnote{Pysynphot: \url{https://pysynphot.readthedocs.io/en/latest/}} \citep{pysynphot+2013} and WD spectra templates \citep{Koester+2010, Tassoul+1990, Tremblay+2009}, the theoretical bolometric luminosities were converted to the HST-WFC3 magnitudes. A selection of the evolutionary tracks are plotted in Figure~\ref{fig:3}. As expected, all the tracks are located along the red side of the CO-WD sequence. The CMD position of the companion star to NGC~6652B is nicely reproduced in both the filter combinations by the lowest available He-WD mass of $0.17~M_\odot$, thus confirming that the identified star is a He-WD. Unfortunately, the available evolutionary tracks do not properly cover the extreme He-WD low mass end (for $M<0.17~M_\odot$).
This hampers a detailed characterization of the uncertainties (see \citet{Cadelano+2019, Cadelano+2020}). However, the main physical properties of the companion star can be safely derived from the best-fit He-WD track mass ($0.17~M_\odot$) and deriving conservative uncertainties exploiting the available grid of higher mass tracks. In doing so the companion star turned out to have a mass of $0.17\pm0.02~M_\odot$, a surface gravity of $\log g = 5.7\pm0.3 \, cm \, s^{-2}$, a surface temperature of $T_{eff}=11500\pm1900 \, K$ and a very young age of only $200\pm100$~Myr. This confirms that the companion to NGC~6652B is a recently formed He-WD. For such He-WD, the proto-WD timescale, i.e. the time spent by the companion star from the Roche-Lobe detachment to the beginning of the cooling phase is expected to last $1.2\pm0.2$~Gyr. Although this value slightly depends also on the NS mass and companion metallicity, it suggests that the mass-transfer phase in this system stopped around 1.4~Gyr ago, then the companion went through a long bloated proto-WD stage in which a significant fraction of the hydrogen in the envelope was burned through stable burning, while the star contracted eventually entering in the WD cooling sequence. 
 
\begin{figure}
    \centering
    \includegraphics[scale=0.55]{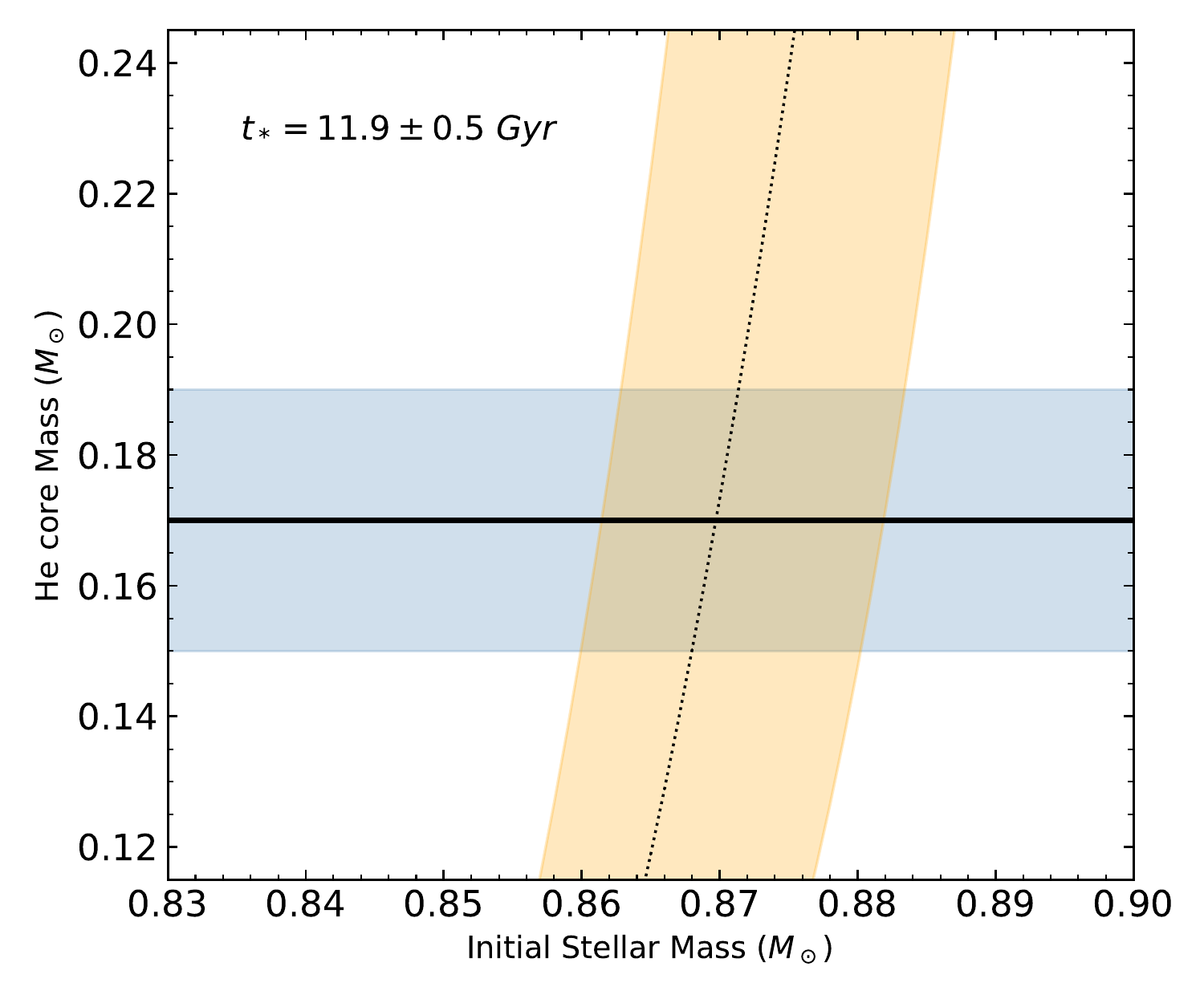}
    \hspace{3pt}
    \includegraphics[scale=0.55]{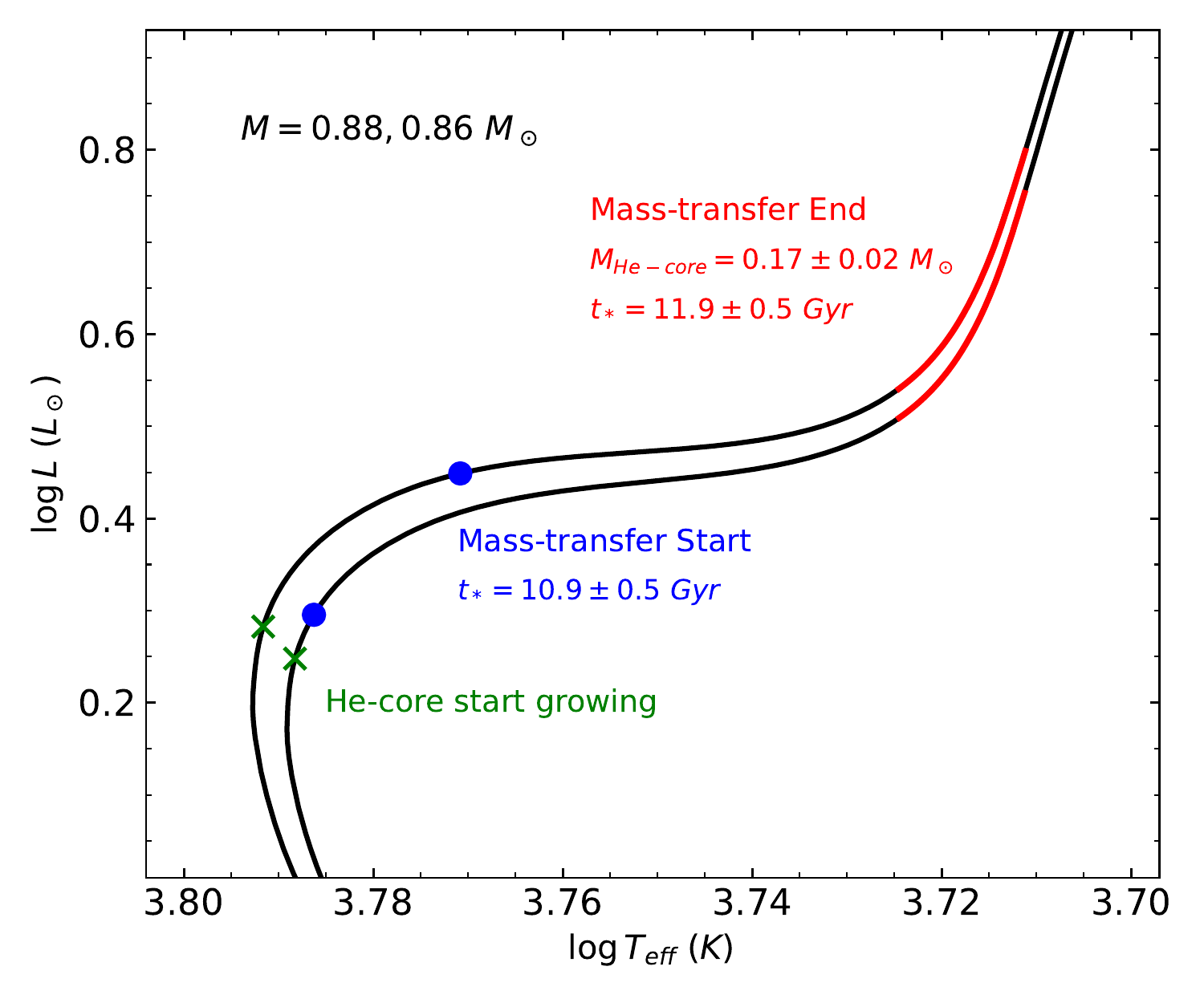}
    \caption{ Left panel: He core mass as a function of the stellar mass as predicted by the evolutionary tracks from PARSEC v2.0 \citep{Costa+2019a, Costa+2019b, Nguyen+2022}. The black dashed curve and the light orange shaded area represent the values for an age of $11.9\pm0.5$ Gyr, corresponding to the cluster age at the epoch of Roche-Lobe detachment. The horizontal line (and blue band) marks the mass of the He-WD companion to NGC~6652B. Right panel: the black curves are the evolutionary track for $0.88 \, M_{\odot}$ and $0.86 \, M_{\odot}$ stars from left to right, respectively. The red portion  of the track highlights the phase where the star had a He core with a mass comparable with the one measured for the companion star.  The blue point  marks the region of the tracks $\sim1$ Gyr before the Roche-lobe detachment when the mass transfer probably started. Finally, the green cross marks the region where the exhausted He-core starts growing.}
    \label{fig:4}
\end{figure}

 To further investigate the evolution of this system, we analyzed the growth rate of the He core of stars with different masses at the cluster metallicity extracted from the PARSEC v2.0 database \citep{Costa+2019a, Costa+2019b, Nguyen+2022}. Assuming the cluster age of $13.25\pm0.5$ Gyr \citep{Dotter+2010}, the sum of the companion cooling age and proto-WD age suggests that this system experienced the end of the mass-transfer phase (i.e. Roche-Lobe detachment) when the cluster was $11.9\pm0.5$ Gyr old. Back then, only stars with masses between $0.86 \, M_{\odot}$ and $0.88 \, M_{\odot}$ had sufficient time to grow a He core with a mass comparable with that of the companion star (see left hand panel of Figure~\ref{fig:4}). This pose a firm constrain of the mass of the progenitor companion star. Stars with mass in the range $0.86 \, M_{\odot}$ and $0.88 \, M_{\odot}$ have developed a $0.17\pm0.02 \, M_{\odot}$ He-core during the very early stages of the red giant branch phase (see right hand panel of Figure~\ref{fig:4}). Assuming that the mass-transfer phase lasted at most $1~Gyr$, then this process started when the companion star evolved to the sub-giant branch stage. 

\subsection{Constrains on the NS mass}

\begin{figure}
    \centering
    \includegraphics[width=18cm]{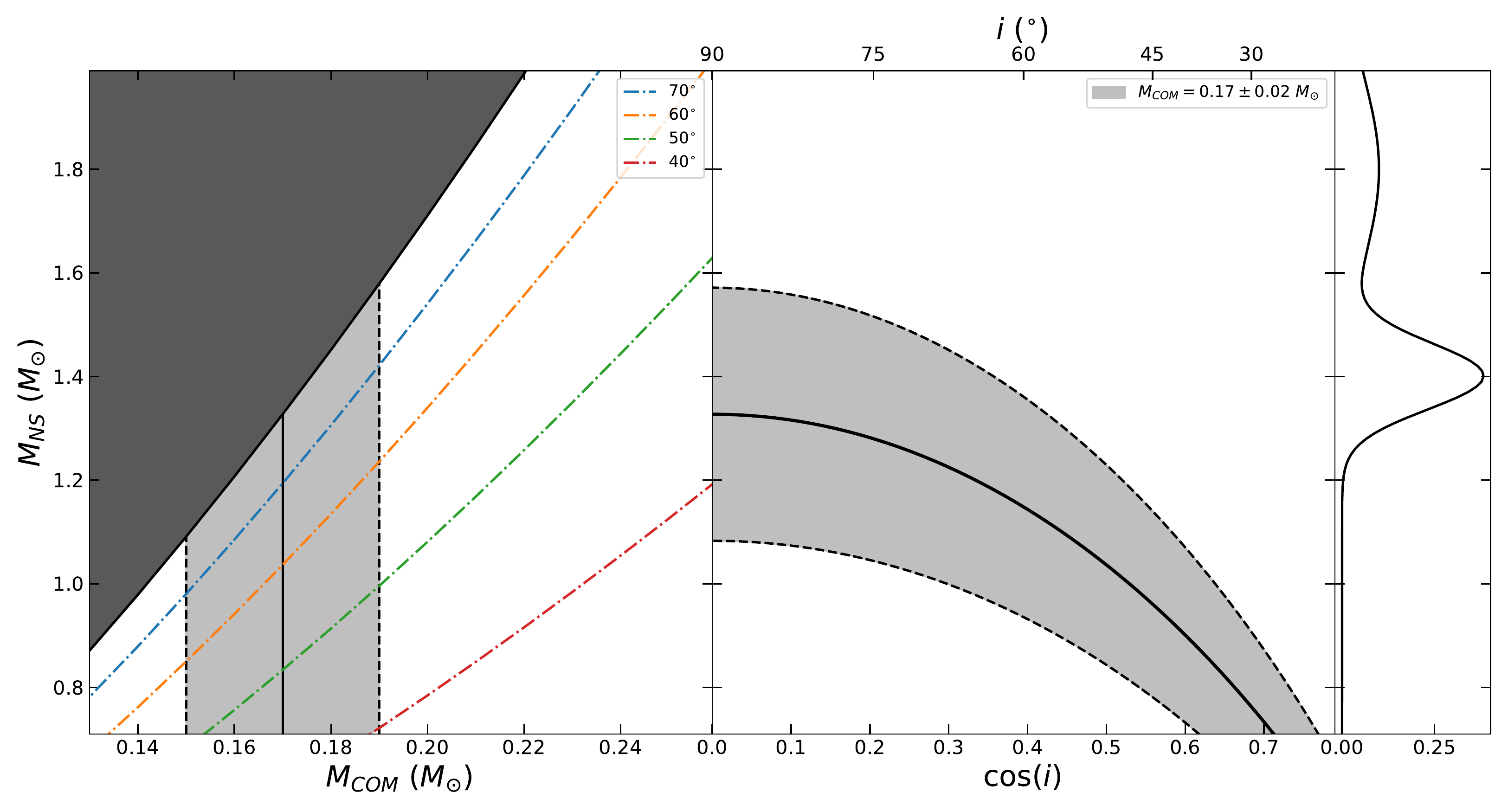}
    \caption{{\it Left-hand panel}: NS mass as a function of the companion star mass. The best-fit value of the companion mass is marked with a solid vertical line, and the corresponding uncertainties are delimited by two dashed lines. The color dotted-dashed curves indicate the NS mass - companion mass relation at different inclination angles, while the shaded area is the region forbidden by the mass function.
    {\it Right-hand panel}: The NS mass as a function of the cosine of the inclination angle for the estimated mass of the companion star. The solid curve are values predicted by the best-fit value of the companion mass, while the light-gray shaded region delimited by the two dashed curves are the values allowed within the companion mass uncertainty. The rightmost panel shows, as a reference, the NS mass distribution empirically derived by \citet{Antoniadis+2016b}.}
    \label{fig:5}
\end{figure}

With the determination of companion star mass, one can combine binary orbital parameters, derived through the radio timing, to constrain the mass of NS from the mass function \citep{Lyne+1998} :

\begin{equation}
	f(m_p, m_c, i) = \frac{4\pi^2}{G} \frac{a^3\sin^3 i}{P^2_b}
	= \frac{m^3_c \sin^3 i}{(m_p + m_c)^2}
	\label{equation2}
\end{equation}

where $m_p$, $m_c$, are the NS and companion mass, respectively, $i$ is the orbital inclination angle, $a$ the projected semi-major axis and $P_{orb}$ the binary period. The above equation contains two unknown quantities, namely the NS mass and the orbital inclination angle. The NS mass as a function of the companion mass and inclination angles are shown in Figure~\ref{fig:5}, where it can be seen that the range of NS masses allowed by the known companion mass excludes the possibility of a high-mass NS. In fact, in the case of an edge-on binary, the NS mass is constrained between $1.1 \, M_{\odot}$ and $1.6 \, M_{\odot}$. Lower inclination angles would result in lower NS mass ranges. In principle, a canonical mass NS of $M_{NS}=1.4~M_\odot$ can be obtained with a large inclination angle of about $i=80-90\degree$. 

\begin{figure}
    \centering
    \includegraphics[width=8cm]{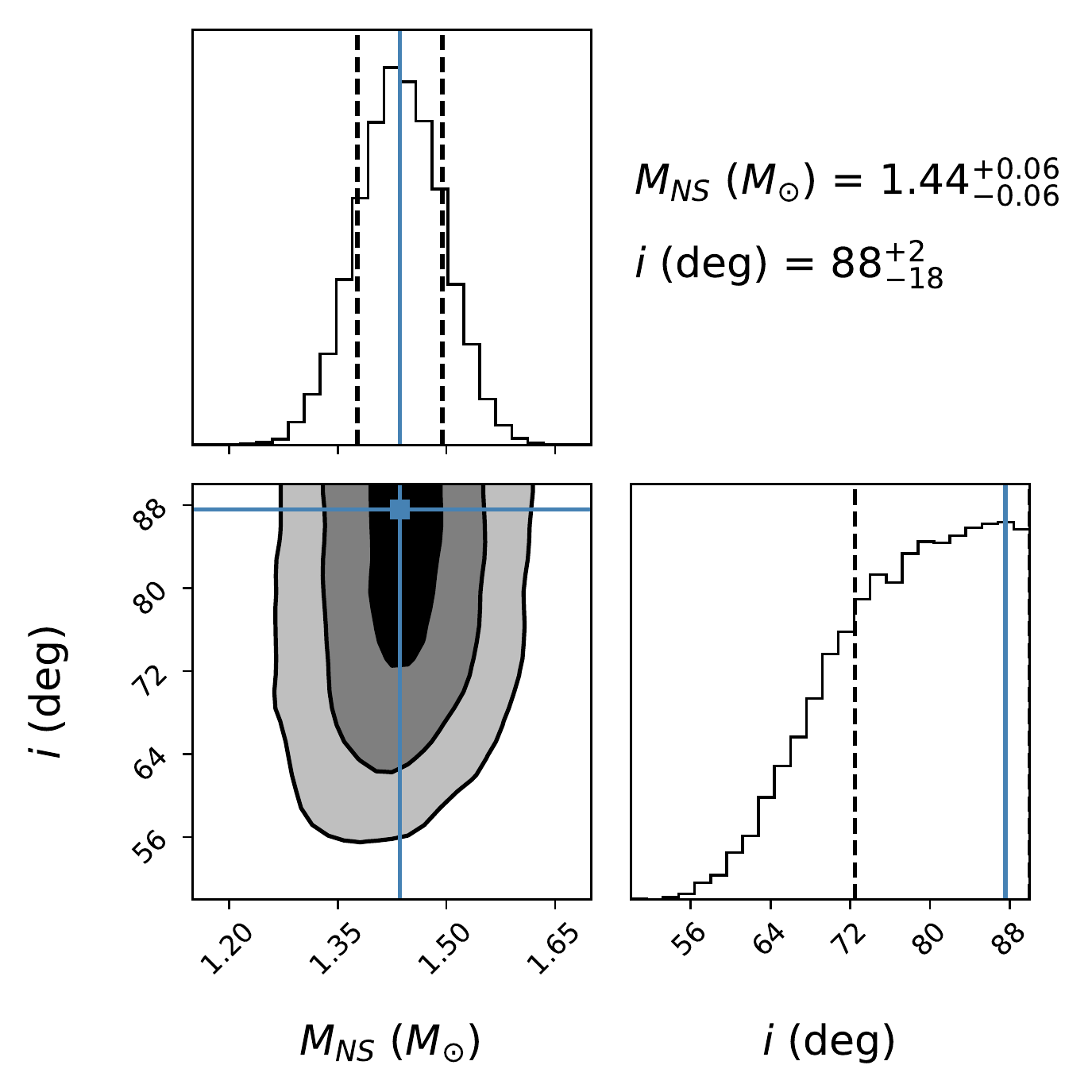}
    \caption{Constraints on the mass of the NS and the inclination angle of NGC 6652B. The 2D panel shows the posterior probability distribution of the two parameters, and the contours are the $1\sigma$, $2\sigma$, and $3\sigma$ confidence levels. The 1D histograms are the marginalized probability distributions of the two parameters, where the solid blue and black dashed lines are the best values and their related uncertainties.}
    \label{fig:6}
\end{figure}

 We also used a Markov Chain Monte Carlo sampler \citep{Foreman-Mackey+2019} to further constrain the mass and inclination angle of the MSP. Following \citet{Cadelano+2020}, we defined a  Gaussian likelihood function to minimize the difference between the middle and right sides of Equation~\ref{equation2}. We assumed a uniform prior on the distribution of $\cos i$ and also a prior on the NS mass distribution following the one derived by  \citet{Antoniadis+2016b}. The posterior distribution is shown in Figure~\ref{fig:6}, and the results based on the $16^{th}$, $50^{th}$, and $84^{th}$ percentiles show that NGC~6652B is likely a NS with a standard mass of $1.44\pm0.06 \, M_{\odot}$ in a binary seen with a large inclination angle. In fact, the probability distribution of the inclination angle reaches the maximum values for edge-on orbits. We therefore assumed that the best
value for the inclination angle corresponds to the maximum in
the probability distribution and its lower uncertainty to the $16^{th}$
percentile: $i \, (deg) =88^{+2}_{-18}$. 

\section{Conclusions}
\label{sec:4}

We used deep, high-resolution near-ultraviolet HST observations to study the binary MSPs in NGC 6652. By using the so-called ``UV-route" approach, we searched for the companions to the binary MSPs in the cluster. At the corresponding position of NGC~6652B we found a blue object located along the red side of the brightest portion of the WD cooling sequence. This is a He-WD, i.e. the exhausted core of an evolving star which lost its envelope likely due to the mass transfer onto the NS. The comparison with binary evolution models revealed that the companion star is a newly formed He-WD with a cooling age of only $200\pm100$~Myr and a mass of $0.17\pm0.02 \, M_{\odot}$. The companion underwent a bloated proto-WD phase which lasted $1.2\pm0.2$~Gyr. Therefore the Roche-Lobe detachment occurred $\sim1.4$ Gyr ago, when the cluster was 11.9~Gyr old. The progenitor star of such a young and low-mass WD was likely a $\sim0.87 \, M_{\odot}$ star which developed a He core with a mass of $0.17 \, M_{\odot}$ during the first stages of the evolution along the red giant branch, before the Roche-Lobe detachment. Therefore the progenitor star lost $\sim0.7 \, M_{\odot}$ during its evolution in the binary system. Assuming that the mass-transfer stage lasted $\lesssim 1$ Gyr, it begun when the progenitor star evolved along the sub-giant branch phase, naturally  increasing its radius due to the standard stellar evolution, eventually filling its Roche-Lobe and starting the mass-transfer onto the NS primary.

Combining together the companion mass here derived and the binary orbital properties derived through radio timing, we found that this system likely hosts a NS with a mass of $1.4 \, M_{\odot}$ seen at an almost edge-on orbit. The NS mass so derived is very similar to the canonical values expected and measured for these objects \citep[e.g.][]{Antoniadis+2016b}. Since the companion lost $\sim0.7 \, M_{\odot}$ during the mass-transfer, the derived NS mass suggests that the mass-transfer phase was highly non-conservative. Future observations will improve the timing precision of this pulsar. Given the almost edge-on orbit predicted for this system, we expect a possible detection of the Shapiro delay in the future. Such a measurement will provide an independent estimate of the companion mass and orbital inclination angle, thus also granting a benchmark for binary evolution models.

{As described by \cite{Gautam+2022}, NGC~6652B is an unusually bright pulsar, with the pulsed component representing only a small amount of the total emission. Such an unusual feature was inferred from the fact that the pulsar appears brighter in interferometric images than in the pulsed profile. The peculiarity of this system was further confirmed by \cite{Zhang+2022}. In fact, their detection of gamma-ray pulsations shows that this MSP is unusually energetic: it is only the third pulsar
in a GC detected in gamma rays, after NGC~6624A \citep{Freire+2011} and M28A \cite{Johnson+2013}. From the orbital period derivative of the pulsar, \cite{Gautam+2022} estimated the intrinsic spin-down of the
pulsar $\dot{P}_{\rm int} <6.65\, \times \,  10^{-20}$, corresponding to a characteristic age $\tau_c>0.43$ Gyr and a gamma-ray efficiency of 0.12. The latter value number indicates that $\dot{P}_{\rm int}$ cannot be much smaller than the derived upper limit, otherwise the gamma-ray efficiency would be unreliably close to 1.0. This, in turns, means that $\tau_c$ cannot be much larger than the derived lower limit. All this suggest that NGC~6652B is an exceptionally young and powerful MSP and the detection of the young WD companion presented in this work is in perfect agreement with this scenario. 
The cooling age here derived is smaller than the characteristic age (assuming $\tau_c \sim 0.43$ Gyr). This is not surprising, because $\tau_c$ is computed assuming that the initial spin period is much smaller than the current spin period, which is unlikely for a MSP with a short current spin period of 1.83 ms. For the much more probable case that the initial spin period was similar to what it is today, then the real age of the pulsar is smaller than suggested by $\tau_c$ and more in agreement with the WD cooling age.} 

This work further confirms the strong and fruitful synergy between the new radio telescopes (such as MeerKat and FAST), and space telescopes (such as HST and JWST). Such a synergy allows a comprehensive view of the properties and evolution of binary MSPs in GCs through the observation of both the NS primaries in the radio bands and the companions in the optical bands.

\vspace{8mm}

This work is part of the project Cosmic-Lab at the Physics and Astronomy Department ``A. Righi'' of the Bologna University (\url{http://www.cosmic-lab.eu/Cosmic-Lab/Home.html}). The research was funded by the MIUR throughout the PRIN-2017 grant awarded to the project Light-on-Dark (PI: Ferraro) through contract PRIN-2017K7REXT. JC acknowledges the support from China Scholarship Council (CSC).


\vspace{5mm}
\facilities{HST(WFC3).}
\software{DAOPHOT\citep{Stetson+1987}, ALLFRAME\citep{Stetson+1994}, Pysynphot\citep{pysynphot+2013}, corner.py\citep{Foreman-Mackey+2016} , emcee\citep{Foreman-Mackey+2019}.
}

\appendix

\section{PSR~J1835-3259A}
\label{sec:app}
\begin{figure}[!htbp]
    \centering
    \includegraphics[width=5.6cm]{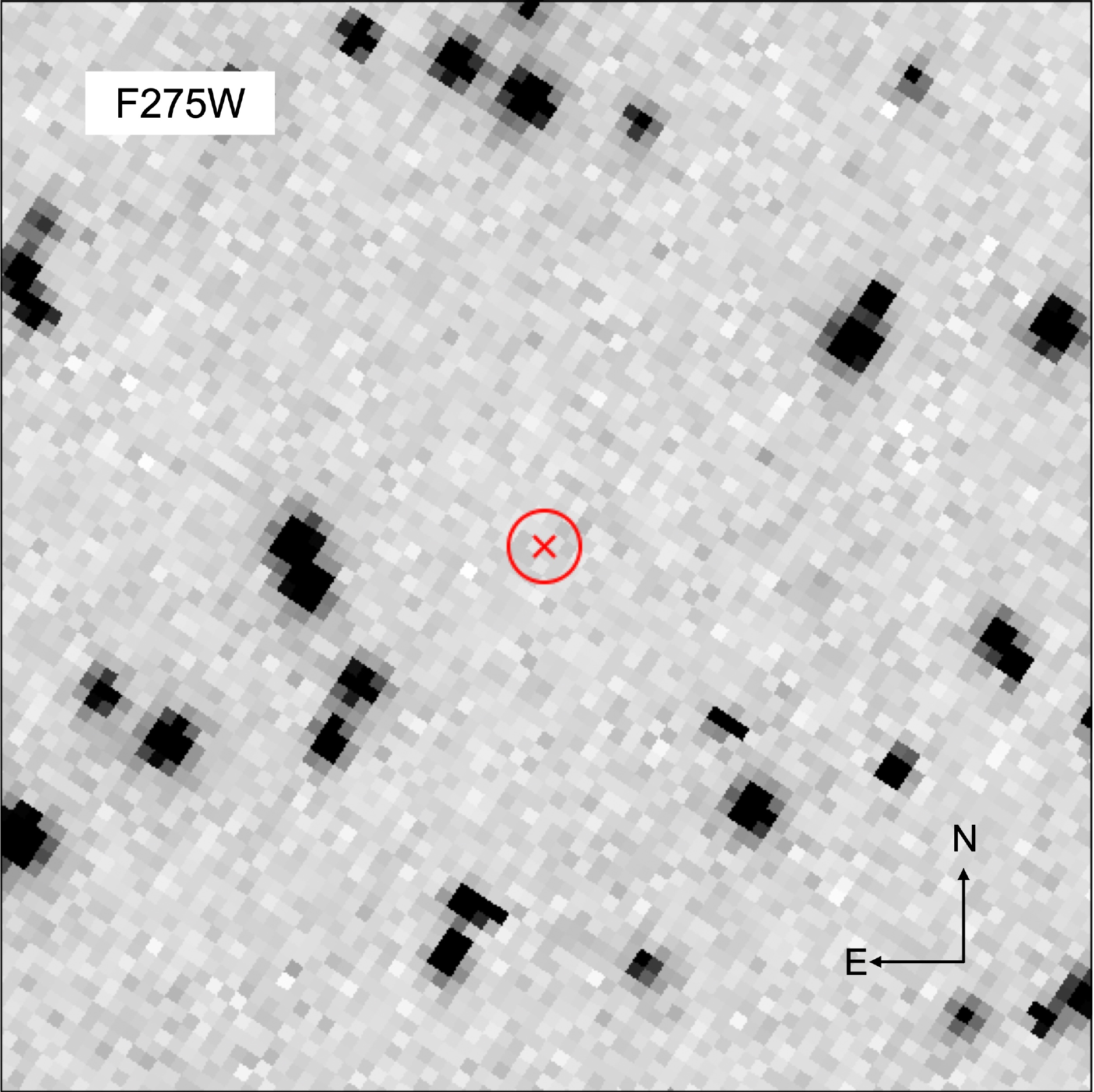}
    \hspace{5pt}
    \includegraphics[width=5.6cm]{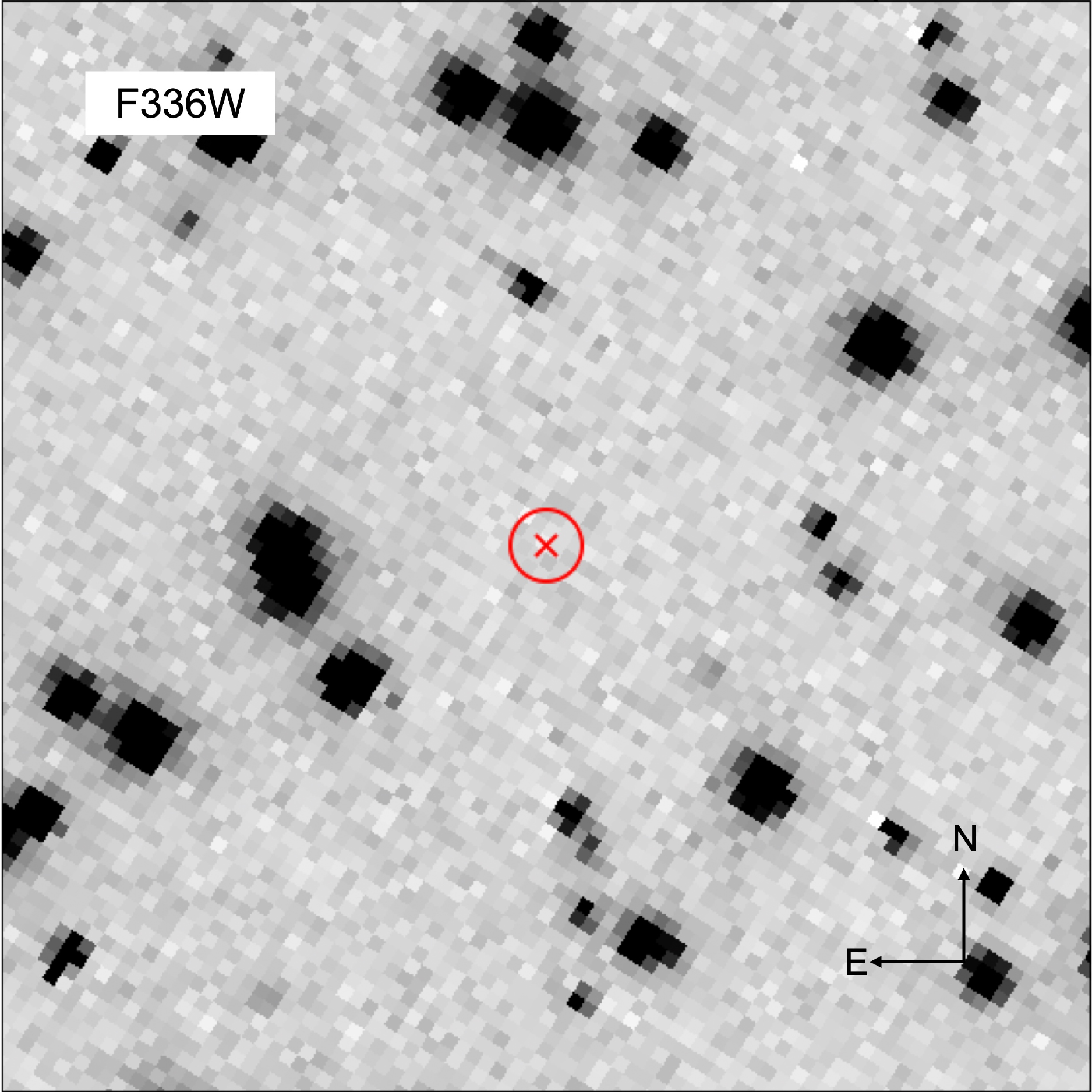}
    \hspace{5pt}
    \includegraphics[width=5.6cm]{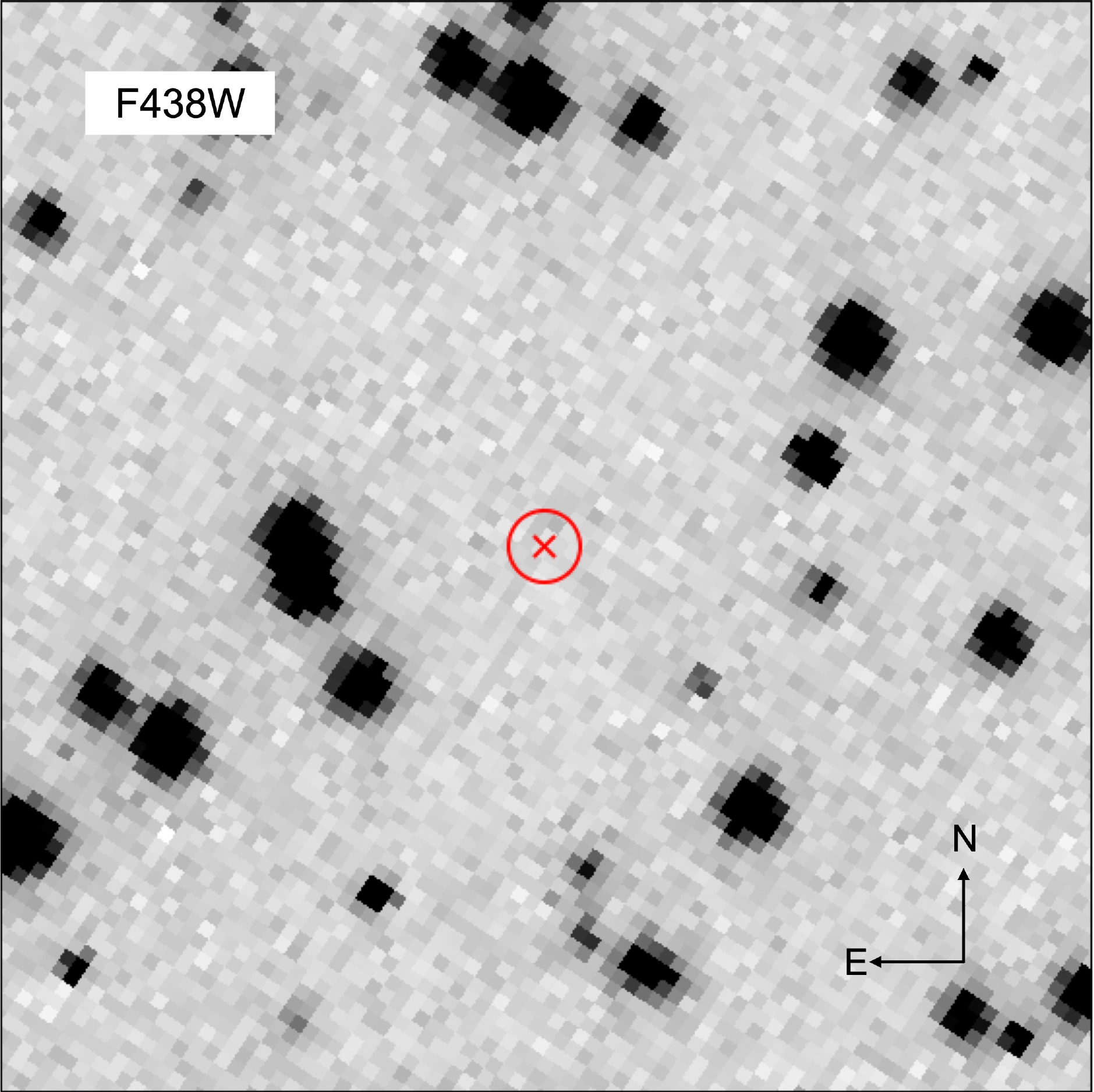}
    \caption{$3\arcsec \times 3\arcsec$ finding chart of the region surrounding NGC~6652A  in the F275W, F336W, and F438W filters. The red circle has a radius of 100 mas ($\sim6\sigma$ uncertainty), while the center of the circle corresponds to the position of the MSP. No stars are detected at the pulsar corresponding position. }
    \label{fig:chartA}
\end{figure}

We carefully investigated the region surrounding the position of PSR~J1835-3259A (NGC~6652A) but no stars are found at its corresponding position in the near-UV and optical images. Figure~\ref{fig:chartA} shows the finding chart in the three WFC3 adopted filters. We also provide the magnitude upper limits of this system derived following \citet{Cadelano+2015a}: $m_{F275W}>25.5$, $m_{F336W}>25.5$ and $m_{F438W}>23.0$.

\end{document}